# On the origins of order.

Jeffrey J. Fredberg

Harvard T.H. Chan School of Public Health
Boston, MA 02115
jfredber@hsph.harvard.edu

12/05/21

Two sentence summary: In his groundbreaking book *The Origins of Order*, the natural philosopher Stuart A. Kauffman proposed the general biological principle that living systems exist in a solid regime near the edge of chaos, and that natural selection achieves and sustains such a poised state. In this brief perspective I propose that, in certain systems at least, this poised state as predicted in the abstract by Kaufmann is realized in the particular by the jammed regime just at the brink of unjamming.

Abstract: A cardinal feature common to embryonic development and tissue reorganization, as well as to wound healing and cancer cell invasion, is collective cellular migration. During collective migratory events the phenomena of cell jamming and unjamming are increasingly recognized, and underlying mechanical, genomic, transcriptional, and signaling events are increasingly coming to light. In this brief perspective I propose a synthesis that brings together for the first time two key concepts. On the one hand, it has been suggested that the unjammed phase of the cellular collective evolved under a selective pressure favoring fluid-like migratory dynamics as would be required so as to accommodate episodes of tissue evolution, development, plasticity, and repair. Being dynamic, such an unjammed phase is expected to be energetically expensive compared with the jammed phase, which evolved under a selective pressure favoring a solid-like homeostatic regime that, by comparison, is non-migratory but energetically economical and mechanically stable. On the other hand, well before the discovery of cell jamming Kauffman proposed the general biological principle that living systems exist in a solid regime near the edge of chaos, and that natural selection achieves and sustains such a poised state. Here I propose that, in certain systems at least, this poised state as predicted in the abstract by Kaufmann is realized in the particular by the jammed regime just at the brink of unjamming.

INTRODUCTION:

Inert disordered collective systems are commonplace. Examples include sand in a pile or grain in a hopper as well as powders, pastes, colloids, slurries, suspensions, and foams. Living disordered collective systems are also familiar, with perhaps the best example being the cellular assembly comprising a confluent multicellular tissue such as the epithelial layer that covers body surfaces and lines the lumen of every hollow organ. Whether inert or living, it is thought that disordered collective systems such as these attain a collective solid-like phase when they jam and attain a fluid-like phase when they unjam. This notion was first suggested by Cates et al.[1], popularized by Liu and Nagel[2], later elaborated by others[3,4] for the case of inert granular materials, and by still others for the case of living multicellular tissues.[5-8] Here I will use a particular model system —tumor cell invasion— as an example by which to generalize these findings and set them into a broad biological context.

What's so special about jammed systems?  For at least three reasons jammed systems have garnered special interest in soft matter science and physical biology.

The first reason that jammed systems warrant special attention is that each constituent particle —whether a grain in sandpile or a cell in a tissue— can become trapped, *i.e.,* caged, by surrounding nearest neighbors and thereby undergo kinetic arrest. When a small force is applied, that force then becomes transmitted in chains from particle-to-particle, or cell-to-cell, and these force chains assemble into a network that can support an applied shear stress if not indefinitely, then for a very long time.[1, 9] By this criterion the material collective comprises an elastic solid. Examples of such solid-like jammed phases include bubbles comprising a stable shaving foam, sand grains comprising a stable sandpile, or cells comprising a stable living tissue. When an applied shear stress exceeds the yield stress, however, then such a collective system can unjam and thus transition from a solid-like phase that is shape-stable to a fluid-like phase that readily flows. The second reason for special attention is that the solid-like jammed phase is trapped far away from thermodynamic equilibrium. System energy is not minimized, therefore, and principles derived from notions of either thermodynamic equilibrium or small departures therefrom, such as theories of rubber elasticity or viscoelasticity, fail to explain the material properties of the solid-like phase or the fluid-like phase or the transition between them. The shortcoming of these theories is attributable to the fact that thermal fluctuations in such systems are far too small to drive microstructural rearrangements of trapped constituent particles over constraining energy barriers that define a rugged energy landscape, thereby trapping the system away from its minimum-energy equilibrium configuration. That is to say, these jammed systems can be regarded as being effectively athermal. Finally, unlike equilibrium thermal systems such as water, which attains a solid phase upon cooling by virtue of a structural transition from disorder to order, —*i.e.,* a crystal— these athermal systems attain a solid-like phase by virtue of caging and resulting kinetic arrest, and thus remain disordered —*i.e.,* amorphous— in the solid-like and fluid-like phases alike. For these three reasons, taken together, jammed systems are of great interest but remain poorly understood.[3]

Jamming in confluent tissues: In the case of confluent tissues, cell jamming and unjamming are now increasingly recognized in the context of collective epithelial cellular migration, which in the healthy tissue is a cardinal feature of tissue development, plasticity, and repair. To cite just a few prominent examples, cell jamming and unjamming have been identified in embryonic development and tissue reorganization in the healthy tissue [10-13], as well as in wound healing [7], cancer cell invasion[14, 15], asthma[5], and idiopathic pulmonary fibrosis[16]. Underlying mechanical, genomic, transcriptional, and signaling events are increasingly coming to light.[15, 17, 18]

In the case of confluent tissues, energy barriers that serve to impede structural rearrangements, and thus can cause the system to jam, are thought to vanish altogether in certain defined circumstances, in which case the system unjams.[19-21] As addressed below, energy barriers that typify the jammed system —which by definition is solid-like, elastic, and static— are not to be confused with frictional energy dissipation, and associated metabolic demands as would be required to propel a migrating unjammed system. [22, 23]

A particular model system: To illustrate the central thesis of this communication, I use recent observations of cancer cell invasion from a tumor spheroid into a surrounding connective tissue matrix, and consider these experimental observations through the lens of a minimal 2-dimensional computational model (Fig. 1).[24] In brief, the model begins with a confluent cluster of cells in a 2D matrix. Using an agent-based approach, each cell expresses capacities for mutual volume exclusion, cell-cell adhesion, cortical tension, elasticity, viscosity, propulsion, and polarization. From an initial state, the system is then allowed to evolve as cells migrate into a surrounding matrix comprising fixed post-like obstructions at a prescribed spatial density, which are taken to represent connective tissue fibers (Fig. 1, insets A,B,C). This minimal model is sufficiently complex to express a remarkably rich variety of phenomenologies, and thus becomes a useful tool to fix certain ideas as generalized below. If such a model were elaborated further, such as by extension to 3-dimensions or by incorporation of cell-based proteolytic digestion of connective tissue, for example, the resulting jamming phase diagram might differ in its details, both large and small. But the central argument put forward below is insensitive to those details. Of course this minimal computational model has numerous controllable variables. But for illustrative purposes we restrict attention to only two key variables that are systematically varied: collagen density and cellular propulsive force (also referred to as cell motility). For each combination the locus of every cell was determined as a function of time.

Three migratory phenotypes: Three migratory phenotypes are observed. First, when cell motility is sufficiently small or collagen density is sufficiently large, only minimal invasion is observed (Fig. 1 inset A). Each cell remains confluent with its immediate neighbors and restricted to a locus close to its original position. This computational simulation reasonably approximates the phenotypic behavior observed experimentally when a spheroid comprising 1000 to 5000 non-tumorigenic epithelial cells (MCF10A) is embedded in collagen at a density of 4 mg/ml. Experimental data indicate that cell shapes are cobblestone in appearance with elongation aspect ratios consistent with cells close to the jamming transition.[25-27] Kang, Ferruzzi *et al*. referred to this situation as the solid-like jammed regime.[24]

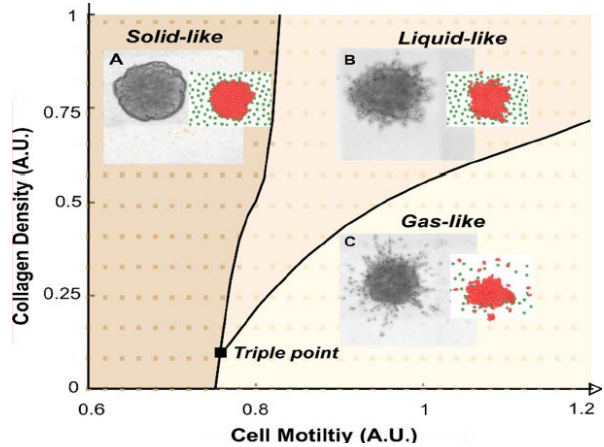

**Fig. 1.** By one computational model, solid-like, liquid-like, and gas-like modes of cell invasion into matrix are observed, as are co-existence lines at which two phases can co-exist, and a triple point at which all three phase can co-exist. Adapted from Kang, Ferruzzi *et al*.,[24]

Second, when cell motility is made sufficiently larger, or collagen density is made sufficiently smaller, there arise continuous invasive tongues that penetrate into the matrix (Fig. 1 inset B). This computational simulation reasonably approximates the cooperative phenotypic behavior observed experimentally when a spheroid of MCF10A cells is embedded in collagen at a density of 2 mg/ml, as well as the behavior that is observed when a cluster of aggressive MDA-MB-231 cells is embedded in collagen at a density of 4 mg/ml; MDA-MB-231 cells are post-metastatic and express mesenchymal markers including high Vimentin and low E-Cadherin. Experimental data indicate that cell shapes at the base of the invasive tongue are systematically elongated and migrate faster than cells in the MCF10A spheroid. Kang, Ferruzzi *et al.* referred to this as the fluid-like unjammed regime.[24] They noted, however, that with increasing penetration into the matrix cells comprising the invasive tongue tend to rejam. For the invading cellular collective, these observations support the interpretation that high density collagen promotes steric hindrance and an associated confinement-induced re-jamming.[28, 29]

Third, when cell motility is sufficiently large but collagen density is sufficiently small, individual cells and small cell clusters are seen to separate from the primary spheroid, scatter, and invade into the matrix. This computational simulation reasonably approximates the phenotypic behavior observed experimentally when a spheroid of the aggressive MDA-MB-231 cells is embedded in collagen at a density of 2 mg/ml. Experimental data show that, depending upon experimental parameters, cell shapes and velocities within the spheroid can indicate either a fluid-like or a solid-like regime, whereas cells scattering into matrix comprise a gas-like regime.[24]

Gradual perturbations beget abrupt changes: How does the transition between these different modes of invasion occur as a function of the degree of matrix confinement? Using graded concentrations of collagen (1 to 4 mg/ml), Kang, Ferruzzi et al. tracked over time the number of single MDA-MB-231 cells that had detached from the continuous primary spheroid, escaped that spheroid, and invaded in a gas-like fashion into the matrix (Fig. 2). The number of such single invading cells or cell clusters was found to be dependent not only upon time but, more importantly, dependent upon collagen centration (Figure 2). On day 0 no cell escape is evident at any collagen concentration; immediately after embedding in collagen, all cells remain within the spheroid. On day 1 a modest level of cell escape becomes evident at lower collagen concentrations (1 and 2 mg/ml) but not at higher concentrations. On day 2 the number of detached invading cells becomes much larger and highly sensitive to collagen concentration. By day 3, remarkably, the number of detached invading cells stabilizes into a striking, almost biphasic, dependence on collagen concentration. The collagen concentration demarking this step-like transition for MDA-MB-231 spheroids falls between 2 and 3 mg/ml. Overall, these findings support the existence of a critical collagen density at which MDAMB-231 cells at the spheroid periphery transition in an almost switch-like fashion between distinct modes of invasion, namely, liquid-like versus gas-like. These events likely depend upon active remodeling of matrix by metalloproteases, cell generated traction forces, and the manner in which these tractions act to align collagen fibers. [30-32]

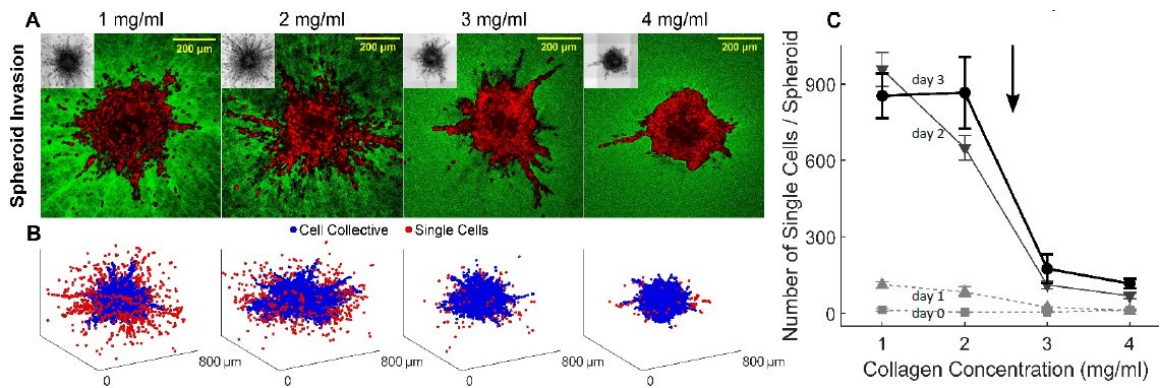

**Fig. 2** A. Invasion of a cancer spheroid comprising MDA-MB-231 cells into collagen matrix at graded degrees of collagen density. B. Cells confluent with the primary spheroid are depicted in blue, and those that have scattered from the spheroid are depicted in red. C. The number of detached single cells or small cell clusters as a function of days of maturation and collagen density. By day 3, remarkably, the number of detached and invading cells stabilizes into a striking, almost biphasic, dependence on collagen concentration. Adapted from Kang, Ferruzzi et al.,[24]

Mapping a hypothetical jamming phase diagram: Computational simulations performed over a range of cell motilities and collagen densities reveal a jamming phase diagram that depicts solid-like, liquid-like, and gas-like regimes (Fig. 1). The boundaries between these regimes define coexistence lines between phases, *i.e.*, conditions for which two phases can co-exist. The boundaries between phases also depict a triple point at which all three phases can co-exist. The transition from solid-like to liquid-like behavior in this non-equilibrium system is reminiscent of transitions that occur in familiar equilibrium systems: the transition from sold-like to liquid-like regimes is reminiscent of melting; the transition from liquid-like to gas-like regimes is reminiscent of evaporation; and the transition from solid-like to gas-like regimes is reminiscent of sublimation. The existence of each of these non-equilibrium transitions and associated coexistence of these non-equilibrium phases has been confirmed experimentally.[24]

Where does physiology happen? In this jamming phase diagram, where does physiology play out? I argue here that certain regions within this phase space are more conducive to essential multicellular physiological functions whereas others are less so. Consider the following gedankenexperiment, beginning with the idea that all living tissue is to some degree heterogeneous in space. Therefore, when the system as a whole is deep in a solid-like regime we might expect to find a few small islands of multiple cells, with each such island being effectively liquid-like, melted, and therefore capable of complex behaviors such as mutual cellular arrangements or collective cellular migration. But these small liquid-like islands are embedded in sea of cells that are solid-like and therefore effectively frozen. As a result, complex behaviors in any one liquid-like melted island are shielded in the sense that they cannot propagate into the surrounding solid-like space, and therefore these islands cannot influence one another. Large perturbations in the local liquid-like island can therefore exert at best only small effects globally, and the system as a whole remains solid-like and frozen. It is perhaps for this reason, as well as others, that the jammed phase has been referred to as being tumor suppressive.[33]

By contrast, when the system as a whole is deep in a liquid-like regime we might expect to find a few small islands of multiple cells that are effectively solid-like and frozen, as it were. Being solid-like, any complex behavior within such an island, such as mutual cellular rearrangements or migration, would have to be highly cooperative. But these small solid-like islands are embedded in a sea of cells that are liquid-like, melted, and, by virtue of viscosity, mechanically dissipative. As such, cooperative behaviors within the small solid-like island are shielded in the sense that they become dissipated and diffused. The behavior in these islands is thus attenuated in the surrounding liquid-like sea, and these islands therefore cannot influence one another. Large perturbations in local solid-like islands can therefore exert at best small effects globally, and the system as a whole remains liquid-like and melted.

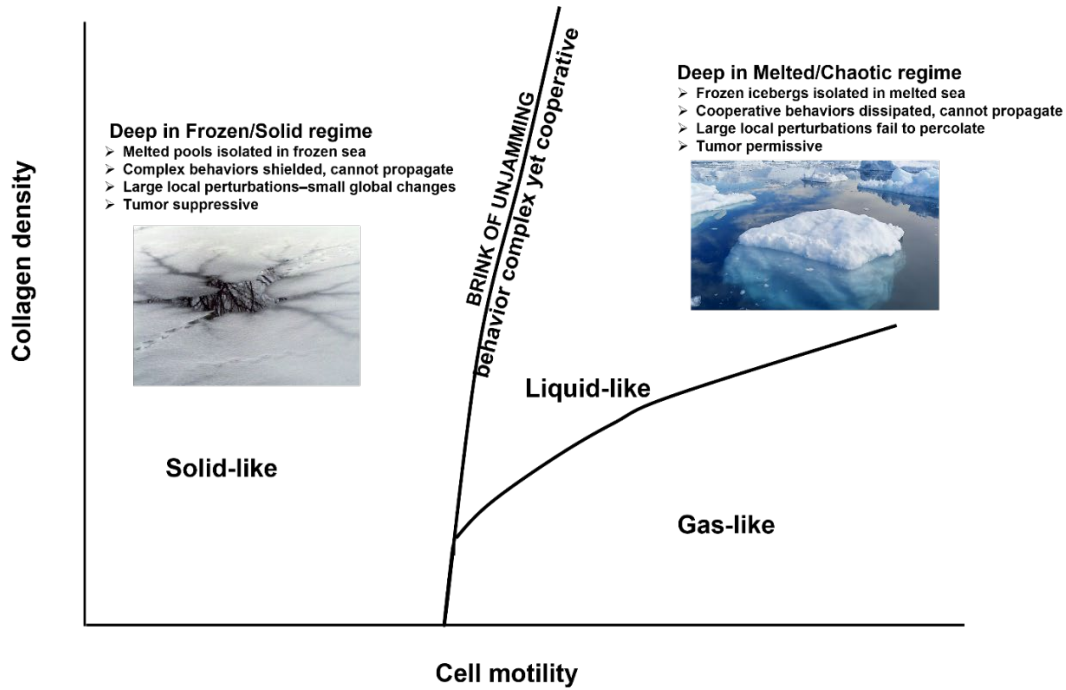

**Fig. 3. In any jamming phase diagram, where does physiology play out?** In the heterogeneous multicellular system, one might expect to find deep in the solid-like regime small islands comprising isolated melted domains. Similarly, one would also expect to find deep in the liquid-like regime small islands of isolated solid-like domains. But in neither situation can the complex and cooperative behaviors that define physiology percolate across the system. Such physiological behaviors include but are not limited to development, remodeling, plasticity, invasion, and wound repair. As such, certain regions within this phase space are more conducive to essential multicellular physiological functions whereas others are far less so. In particular, being poised in the solid-like regime but just at the brink of unjamming allows for the existence —as well as the fine control— of collective cellular behaviors that are not only dynamic and complex but also cooperative and system-spanning. Dysregulation of that control has been implicated in pathologies including cancer cell invasion[14, 15], asthma[5], and idiopathic pulmonary fibrosis[16].

We now consider in particular a special domain, one that is neither deep in the liquid-like regime nor deep in the solid-like regime, but rather is poised in the solid-like regime but just at the brink of unjamming (Fig. 3). For three reasons this regime is singular. First, being poised at the brink of unjamming, this regime allows for collective cellular behaviors that are not only dynamic and complex but also cooperative. Second, this regime is also capable of percolating across space so as to span the physical boundaries of the system. For example, being poised at the brink of unjamming allows for system-wide cooperative dynamical events such as tissue remodeling, wound repair, development, branching, and growth. By contrast, deep within the solid-like regime, as well as deep within the fluid-like regime, for the reasons described above events such as these become increasingly implausible. Finally, being poised at the brink of unjamming allows for rather potent and sensitive mechanical modulation of these processes. By shifting only slightly more deeply into the solid-like regime, for example, dynamic events would become progressively and dramatically slowed or, in the limit, stabilized altogether. Conversely, by shifting only slightly away from the solid-like regime and toward the liquid-like regime dynamic events would become increasing possible and accelerated. Moreover, being thus poised at the edge of unjamming, a very small but well-chosen modulation in certain state variables may alter system-wide behavior rather dramatically, thereby affording a sensitive mechanism for tuning and control of system-wide behavior.

For these reasons, taken together, I propose that in a variety of systems and circumstances, well beyond those illustrated here, physiology happens, and can only happen, in the solid-like regime just at the brink of unjamming. Of course, if this special regime were to become dysregulated it would allow for pathological events such as asthmatic remodeling of the airway wall, idiopathic pulmonary fibrosis, and cancer cell invasion and scattering. [16, 34, 24]

Jamming or unjamming first? In evolution of early multicellular organisms, or in embryonic tissue development, which comes first, jamming or unjamming? Here we turn again to Kauffman, who in 2002 introduced the concept he called the *adjacent possible*.[35] The theory of the adjacent possible proposes that biological systems are able to morph into more complex systems by making incremental, relatively less energy consuming changes in their make up. That is to say, evolution advances mainly by cobbling together available resources to new uses. Based upon this idea, I reason as follows. Being highly dynamic and requiring tissue plasticity and remodeling, the earliest multicellular organisms, as well the developing embryonic tissue, are reasoned to be unjammed but energetically expensive. Little data is available about energy metabolism associated with unjamming. We do know, however, that at the leading unjammed edge of an advancing epithelial layer the cytoplasmic redox ratio becomes progressively smaller, the NADH lifetime becomes progressively shorter, and the mitochondrial membrane potential and glucose uptake become progressively larger.[22] While short of being definitive, these observations are suggestive that the unjammed migratory phase is energetically expensive compared with the jammed phase. To the extent that this might be so, the jammed phase, logically, would represent an obvious adjacent possible with respect to the unjammed phase. In tissue development as in evolution, the unjammed phase, which is dynamic but energetically expensive, is reasoned to precede the jammed phase which follows, which is homeostatic but energetically economical.

What about EMT? Mature tissues are characterized by loss of cell motility, which in turn stabilizes the tissue mechanically and decreases energy metabolism. These features, taken together, suggest that the jamming transition and emergence of the associated solid-like phase may be among the final stages of tissue maturation. But in order for any pathology that dedifferentiates or deregulates epithelial cells, including carcinomas, to become a systemic disease ––as opposed to a localized aberration–– a fluid-like phase allowing for epithelial motility must be restored. In that connection, the unjamming transition is distinct from and not to be confused with the epithelial-to-mesenchymal transition (EMT).[36] Either provides a gateway to epithelial motility. We know, moreover, that the unjamming transition can occur in the absence of EMT, whereas the converse remains unclear.[36] It remains unclear, as well, how unjamming and EMT might work independently, sequentially, or cooperatively to effect morphogenesis, wound repair, and tissue remodeling, as well as fibrosis, cancer invasion, and metastasis.

The origins of order: On the one hand, it has been suggested that the unjammed phase of the cellular collective evolved under a selective pressure favoring fluid-like migratory dynamics as would be required so as to accommodate dynamics associated with episodes of tissue development, plasticity, and repair. This unjammed phase, being dynamic, is reasoned to be energetically expensive compared with the jammed phase, which evolved under a selective pressure favoring a solid-like homeostatic regime that, by comparison, is non-migratory but energetically economical and mechanically stable.[22] On the other hand, well before the discovery of cell jamming Kauffman proposed the general biological principle that living systems exist in a solid regime near the edge of chaos, and that natural selection achieves and sustains such a poised state.[37] Here I propose that, in certain systems at least, this poised state as predicted based upon abstract considerations of complex Boolean networks by Kaufmann[37] is realized in the particular form of the jammed regime of the multicellular system just at the brink of unjamming. A similar situation may be recapitulated at a subcellular scale of organization. [38,39]

**Acknowledgments**: The author thanks Josef Käs for his helpful comments. This work was supported by NIH grant number 1R01HL148152